\documentclass[preprint,12pt]{elsarticle}
\usepackage{amsmath}
\usepackage{amsfonts}
\usepackage{amssymb}
\usepackage[utf8]{inputenc}
\usepackage[T2A]{fontenc}
\usepackage{bm}

\begin{document}

\title{Input of the Coulomb law modification to the Lamb shift of the hydrogen atom} 

\author{A.A.~Eremko, L.S.~Brizhik, V.M.~Loktev}

\begin{abstract}
	 Radiative corrections which remove accidental degeneracy in the spectrum of the relativistic hydrogen atom and lead to the modification of the Coulomb law, are calculated within the novel approach, based on the exact solution of the Dirac equation with the Coulomb potential. The energy spectrum of the hydrogen atom is obtained with account of these corrections and the  Lamb shift is calculated for the lowest energy states.
\end{abstract}

\maketitle 

\vspace{10mm}


\vspace{10mm}

\section{Introduction}\label{Intro}

The most adequate and complete description of the electron states in a nucleus field is provided by the Dirac equation (DE) with the Coulomb potential \cite{Dirac1,Dirac}. The essential consequence  of the relativistic DE is the natural appearance of the electron spin. It has been shown that there exists the fine-structure splitting of the hydrogen-like energy levels which originates from the spin–orbit interaction. The latter removes the orbital degeneracy in the non-relativistic treatment of the problem within the Schr\"{o}dinger equation. But it turns out that it removes degeneracy not completely. For example, according to the DE, the $ 2S_{1/2} $ and $ 2P_{1/2} $ states have the same energy. This so called accidental degeneracy arises from the specific features of the Coulomb potential and is explained by the existence of several, namely Dirac, Johnson-Lippmann and Brizhik-Eremko-Loktev spinor invariants  \cite{BEL-gen}, whose commutators are non-zero.

In 1947 W. Lamb reported an unexpected peculiarity in the fine structure of atomic hydrogen: a separation of the $ 2S_{1/2} $ and $ 2P_{1/2} $ levels, known now  as the Lamb shift. It indicates that the Coulomb law fails at short distance scales. This discovery was a stimulus for the modern development of quantum electrodynamics (QED). Explanation of the Lamb shift as well as of the anomalous electron $ g $-factor are, historically, the major advancements of QED.

Within the covariant formalism of the QED based on the perturbative expansion of the $ S $-matrix,  an electron is studied with account of the influence on it of two fields, one of which is the self-radiation field (electron-photon interaction), and the second one is external electromagnetic field. The former interaction leads to the appearance of radiation corrections (RCs) to the electron interaction with an external field, while the external central potential results in the appearance of the bound electron states with the discrete energies affected by the RCs. It has been shown that the covariant formalism can be generalized, in principle, for the bound states as well, although  calculation of the higher order Feynman diagrams is very cumbersome. Usually, the perturbation theory is used with the description of bound states by solutions of the Schr\"odinger equation. Lamb shift calculations within such a scheme are described in many textbooks on QED, e.g., \cite{Bethe},  and  in the review papers. Nevertheless, it is important that namely the DE provides correct transition to the non-relativistic limit. Taking into account the full set of the DE invariants, the complete set of relativistic corrections in the Schr\"odinger equation has been calculated in  \cite{BEL-2D-el,BEL-nonrel}. Thus, it is worth to calculate the Lamb shift with account of these facts.

In the present paper we calculate the Lamb shift within the conventional in quantum mechanics Hamiltonian description, applied to the DE.  Solving exactly  the DE with the Coulomb potential, the relativistic Hamiltonian of the hydrogen atom is obtained with account of all known RCs, which are calculated using DE exact solution.  The factors decreasing the symmetry and excluding additional invariants from the set of the integrals of motion, remove the energy degeneracy. One of such factors is the deviation of the electric field, which acts on the electron in an atom, from the Coulomb law. The RCs due to interaction of an electron with external fields, are summarized in the next Section.

\section{Radiative corrections to the Coulomb potential } \label{RadCorr}

 The RCs have been considered in the textbook of Bogolyubov and Shirkov \cite{Bogolyubov} (see also \cite{RelQuant}). The RCs include renormalization of the photon and spinor (electron-positron) fields due to the interaction between them, and the vacuum polarization by the electric field of an external charge. To be more exact, the RC essentially arise from the following main factors:\\
-- creation and annihilation of virtual electron-posi\-tron pairs accompanying photon propagation which can be taken into account by the photon polarization (PP) propagator; \\
-- creation and annihilation of virtual photons by an electron which contributes to electron self-energy (SE); \\ 
-- dependence of the vacuum energy on the external constant magnetic and electric fields which leads to electric and magnetic polarization of the vacuum (for instance, by the electric field of a point charge).

To calculate the Lamb shift, we consider below explicit expressions of these RC.
\vspace{5mm}

\subsection{\bf{Effective QED potential from photon polarization}} \label{PhotPol}

The contribution associated with the PP effect, is extracted after employing some techniques within the QED $ \hat{S} $-matrix formalism. In the case of a point charge $ Z e $  (e.g., charge of a nucleus with $Z$ protons) the electric field is given by the Coulomb scalar potential $ \varphi_{C}(r) $. With account of the polarization RC, this field is replaced by "the effective field" \cite{RelQuant,Akhiezer,Gusynin}
\begin{equation}
\label{A0e}
\varphi^{(eff)} = \varphi_{C} + \frac{1}{4\pi} \mathcal{P} \mathcal{D} \varphi_{C}
\end{equation} 
where $ \mathcal{P} $ is the PP operator and $ \mathcal{D} $ is the photon propagator which can be taken in the zero approximation. The second term in this expression $ (1/4\pi) \mathcal{P} \mathcal{D} \varphi_{C} \equiv \delta \varphi_{PP} $ represents the corresponding PP correction of the Coulomb potential. In the coordinate space for a hydrogen atom ($Z=1$), which we will consider below, it has the form  \cite{RelQuant,Akhiezer}
\begin{equation}
\label{Coul_pert} 
\delta \varphi_{PP} (r) = \frac{2\alpha}{3\pi} \frac{e}{r} \int_{1}^{\infty} e^{-\frac{2mcr}{\hbar} \zeta} \left( 1 + \frac{1}{2\zeta^{2}} \right) \frac{\sqrt{\zeta^{2} - 1}}{\zeta^{2}} d\zeta 
\end{equation} 
where $ e > 0 $ is the elementary charge (respectively, electron charge is $- e $) and $\alpha=e^2/\hbar c \approx 1/137 $ is Sommerfeld fine structure constant. 
The potential (\ref{Coul_pert}) is named after E.A. Uehling, who was the first to calculate it in 1935 for a point charge \cite{Uehling}. For the two limiting cases this correction can be represented as \cite{RelQuant,Akhiezer} (see also \cite{Frolov})
\begin{equation}
\label{PP}
\delta \varphi_{PP} (r) = \left\lbrace \begin{array}{c}
\frac{2\alpha}{3\pi} \left( \ln \frac{\lambda_{C}}{r} - C - \frac{5}{6} \right) \frac{e}{r} \quad {\rm {at}} \quad r \ll \lambda_{C} , \\
\frac{\alpha}{4\sqrt{\pi}} e^{-2r/\lambda_{C}} \left( \frac{\lambda_{C}}{r} \right)^{3/2} \frac{e}{r} \quad {\rm {at}} \quad r \gg \lambda_{C}
\end{array} \right. .
\end{equation} 
Here  $\lambda _c=h/mc $ is the Compton wavelength which plays the role of a characteristic length for the spacial behaviour of the PP corrections which exponential decreases at $ r > \lambda _c $.

\vspace{5mm}

\subsection{\bf{Effective QED potential from the electron self-energy}} \label{SE}

In QED, the interaction of a free electron with an external potential source is described by the term $ -e \gamma^{\mu} A_{\mu}^{(e)} (q)$ where $ q = p_{2} - p_{1} $ is the exchange of a four-momentum.  It contains the factor $ -e \gamma^{\mu} $ associated with each space-time point (vertex)  with $ \gamma^{\mu} \, ( \mu=0,1,2,3) $ being the Dirac matrices. The corresponding $ S $-matrix expansion can be combined through the substitution
\[
\gamma^{\mu} \rightarrow \Gamma^{\mu} = \gamma^{\mu} + \Lambda^{\mu}_{R} (p_{1},p_{2}) 
\] 
where 
\begin{equation}
	\label{vert-corr}
\Lambda^{\mu}_{R} (q) = \gamma^{\mu} F_{1} (q^{2} ) + i\frac{\sigma^{\mu \nu} q_{\nu}}{2mc}  F_{2} (q^{2} ), \, \sigma^{\mu \nu} = \frac{1}{2} \left[\gamma^{\mu} ,\gamma^{\nu} \right], 
\end{equation}
is the regularized (physical) vertex-correction function \cite{RelQuant} which gives radiative self-energy correction  into the effective interaction of an electron with the external field $ -e \Gamma^{\mu} A_{\mu}^{(ext)} (q)$. In (\ref{vert-corr}) functions $ F_{1} $ and $ F_{2} $ are known as the electric and magnetic form factors, respectively. 

Using the fact that the effective interaction is sandwiched between the free-electron (on-mass-shell) field operators $ \bar{u}_{p_{1}} $ and $ u_{p_{2}} $, the field of external point charge $ A_{0}^{(ext)} = \varphi_{C} $ generates a vertex-correction potential $ \delta V_{SE}(\mathbf{q}) = - e \delta \varphi_{SE}(\mathbf{q}) $ of the form 
\[
\delta \varphi_{SE}(\mathbf{q}) = \Lambda^{0}_{R} (\mathbf{q}) \varphi_{C} (\mathbf{q}) = \frac{4\pi e}{\mathbf{q}^{2}} \Lambda^{0}_{R} (\mathbf{q})  
\]
which is the sum of the electric and magnetic corrections to the Coulomb potential, $ \delta \varphi_{SE}(\mathbf{r}) = \delta \varphi_{elec}(\mathbf{r}) + \delta \varphi_{magn}(\mathbf{r}) $, corresponding to the respective form factors. 

Since the vertex function depends on the momentum transfer $ \mathbf{q} $ only,  it conveniently yields the local potential in the real space. In the coordinate space the electric form factor gives the scalar potential correction 
\begin{equation}
\label{deltafi_elec} 
\delta \varphi_{elec} \left( r \right) = - \frac{e_{e}\alpha }{ 2\pi r}  \times 
\end{equation}
\[
\int_{1}^{\infty} d\zeta \frac{ -3\zeta^{2} + 2 + \left( 2\zeta^{2} - 1 \right) \ln \frac{4^{2}m^{2}}{\lambda^{2}} \left( \zeta^{2} - 1 \right) }{ \zeta^{2} \sqrt{\zeta^{2} - 1} }  e^{-\frac{2\zeta r}{\lambda_{C}}}.  
\]

The magnetic form factor leads to the correction in the form 
\begin{equation}
\label{deltafi_magn} 
\delta\varphi_{mag} \left( \mathbf{r} \right) =  
\frac{ e \alpha \lambda_{C}}{4\pi } \frac{d\phi \left(r \right)}{dr}
\bm{\hat{\Gamma}} \cdot \mathbf{e}_{\mathbf{r}}, 
\end{equation} 
where 
\begin{equation}
	\label{fi}  
\phi \left(r \right) = \frac{1}{r} \left( 1 - \int_{1}^{\infty} e^{-2r\zeta/\lambda_{C}} \frac{d\zeta}{\zeta^{2} \sqrt{\zeta^{2} - 1}} \right),
\end{equation}
\[\quad 
\bm{\hat{\Gamma}} = \left( \begin{array}{cc}
	0 & - i \bm{\hat{\sigma}} \\
	i \bm{\hat{\sigma}} & 0 
\end{array} \right), \; \mathbf{e}_{\mathbf{r}} = \frac{\mathbf{r}}{r}
\]  
and $\hat{\sigma}_{\mathbf{r}} = \bm{\hat{\sigma}} \cdot \mathbf{e}_{\mathbf{r}} $, $ \hat{\sigma}_{j} $ ($ j=x,y,z $) are the Pauli matrices
\begin{equation}
\label{PM}
\hat{\sigma}_{x} = \left( 
\begin{array}{cc}
0 & 1 \\ 1 & 0
\end{array} \right), \, \hat{\sigma}_{y} = \left( 
\begin{array}{cc}
0 & -i \\ i & 0
\end{array} \right), \, \hat{\sigma}_{z} = \left( 
\begin{array}{cc}
1 & 0 \\ 0 & -1
\end{array} \right).
\end{equation}
These self-energy effective potentials were first derived with respect to a point nucleus (Coulomb potential) \cite{Flambaum,Ginges}. 

It follows from Eqs. (\ref{deltafi_magn}) and (\ref{fi}) that similar to the PP corrections, the Compton wavelength $\lambda_{C}$ here also plays the role of the RC spacial damping. 

\vspace{5mm}

\subsection{\bf{Effective QED potential from the vacuum polarization }} \label{Vac-Pol}

Quantization of the electron-positron field leads to the appearance of an infinite constant $ E_{vac} = - \sum_{\mathbf{p},\sigma} E_{\mathbf{p}}^{(-)} $, where the values $ -E_{\mathbf{p}}^{(-)} $ are negative eigenvalues of the DE. This constant is interpreted as the vacuum energy, from which the energies of the spinor field excitations are counted. In the presence of a constant electromagnetic field, the energies $ E_{\mathbf{p}}^{(-)} $ are shifted, which results in the dependence of the space properties on the field and changes vacuum electromagnetic field equations. This transforms linear Maxwell equations into nonlinear equations and leads to the observable effects, such as light scattering by light or by an external field. W. Heisenberg and H. Euler \cite{HeisEul} were the first to study this phenomenon within the Lagrange density formalism (see also \cite{Peskin}). As a result, the Maxwell equations in vacuum take the form of the macroscopic equations for the field in a material medium
\begin{equation}
\label{Maxw} 
\begin{array}{c} 
\bm{\nabla} \times \bm{\mathcal{E}} = - \frac{1}{c} \frac{\partial \bm{\mathcal{B}}}{\partial t} , \quad \bm{\nabla} \cdot \bm{\mathcal{B}} = 0 , \\ 
\bm{\nabla} \times \left( \bm{\mathcal{B}} - 4\pi \bm{\mathcal{M}} \right) = \frac{\partial}{\partial t} \left( \bm{\mathcal{E}} + 4\pi \bm{\mathcal{P}} \right) , \\
\quad \bm{\nabla} \cdot \left( \bm{\mathcal{E}} + 4\pi \bm{\mathcal{P}} \right) = 0, 
\end{array} 
\end{equation} 
where  
\[
\bm{\mathcal{P}} = \frac{\partial \Delta L}{\partial \bm{\mathcal{E}}} , \quad \bm{\mathcal{M}} = \frac{\partial \Delta L}{\partial \bm{\mathcal{B}}} 
\]
are the electric and magnetic vacuum polarization vectors, correspondingly \cite{RelQuant,Akhiezer}. Here $ \Delta L $ is the change of the Lagrange functional which takes into account the vacuum energy. In the case $ \bm{\mathcal{B}} = 0 $ the vector $ \bm{\mathcal{P}} $ is given by the following expressions:
\[
 \bm{\mathcal{P}} = \frac{\partial \Delta L}{\partial \bm{\mathcal{E}}} = \frac{\hbar^{3} \alpha^{2}}{90 \pi^{2} m^{4}c^{5}} \mathcal{E}^{2} \bm{\mathcal{E}} .
\]

For the central symmetric field one has $ \bm{\mathcal{E}} = \mathcal{E}(r) \mathbf{e}_{\mathbf{r}} $ ($ \mathbf{e}_{\mathbf{r}} = \mathbf{r}/r $) and from the Maxwell equation $ \bm{\nabla} \cdot \left( \bm{\mathcal{E}} + 4\pi \bm{\mathcal{P}} \right) = 0 $ it follows that 
\begin{equation}
\label{electric} 
\mathcal{E} + \frac{2\hbar^{3} \alpha^{2}}{45 \pi m^{4}c^{5}} \mathcal{E}^{3} = \frac{e}{r^{2}} . 
\end{equation}
The integration constant $ e $ in expression (\ref{electric}) is determined from the condition that at $ \alpha \rightarrow 0 $ the field should coincide with the Coulomb field of the elementary charge $ e$. For the case of a nucleus charge $Ze$ the problem can be easily generalized.

Assuming that the point charge field differs little from the Coulomb law $ \mathcal{E}_{0}(r) = e/r^{2} $ due to the smallness of the coefficient $ \alpha $, we can search the approximate solution of non-linear equation (\ref{electric}) in the form of the expansion $ \mathcal{E} = \mathcal{E}_{0} + \alpha^{2} \mathcal{E}_{1} + \ldots $. For the first order correction  of the electric field  we obtain  
\[
\mathcal{E}_{1} = - \frac{2\hbar^{3} }{45 \pi m^{4}c^{5}} \frac{e^{3}}{r^{6}} \left( 1 + \frac{2\hbar^{3} \alpha^{2}}{15 \pi m^{4}c^{5}} \frac{e^{2}}{r^{4}} \right) ^{-1}. 
\]
It is useful to introduce the constant $r_0$  
\begin{equation}
\label{r_0} 
r_{0} = \left( \frac{2 \alpha^{3}}{15 \pi } \right)^{1/4} \lambda_{C} \ll \lambda_{C}
\end{equation} 
which has the dimensionality of the length.
Then the electric field in the first order approximation is given by the expression
\begin{equation}
		\label{Er}
\mathcal{E}(r) = \frac{e}{r^{2}} +\Delta \mathcal{E}_{vac}(r), 
\end{equation} 
\begin{equation}
	\label{dvac} 
 \Delta \mathcal{E}_{vac}(r) = - \frac{e}{3} \frac{r_{0}^{4}}{r^{2} \left( r^{4} + r_{0}^{4}\right) }.
\end{equation} 

For the given electric field the scalar potential is determined by integrating the equality $ d\varphi /dr = - \mathcal{E}(r) = - \left( \mathcal{E}_{0}(r) + \Delta \mathcal{E}_{vac}(r) \right) $, and, therefore,
\[
\varphi (r) = \varphi_{0} (r) + \Delta \varphi_{vac} (r), \quad \varphi_{0} (r) = \frac{e}{r}, \] 
\[ 
\frac{d (\delta \varphi_{vac} (r))}{dr} = \frac{e}{3} \frac{r_{0}^{4}}{r^{2} \left( r^{4} + r_{0}^{4}\right) }.
\] 
Integrating this equation, we get the vacuum polarization correction $ \delta\varphi_{vac} (r) $ to the scalar potential of a point charge 
\begin{equation}
\label{deltafi_vac} 
\delta \varphi_{vac}(r)  =  - \frac{e}{3} \left( \frac{1}{r} + F(r) - \frac{\pi}{2\sqrt{2} r_{0}} \right)   
\end{equation}
where 
\[
F(r) = \int \frac{r^{2}dr}{ r^{4} + r_{0}^{4}} = 
 \frac{1}{2\sqrt{2} r_{0}} \left[ \frac{1}{2}\ln \frac{r^{2} - \sqrt{2} r_{0} r + r_{0}^{2}}{r^{2} + \sqrt{2} r_{0} r + r_{0}^{2}} + \right. 
\]
\begin{equation}
\label{F(r)}  
\left.  + \arctan \left( \frac{\sqrt{2}r}{r_{0}} - 1 \right) + \arctan \left( \frac{\sqrt{2}r}{r_{0}} + 1 \right) \right] .  
\end{equation} 
Here the integration constant $ C = -\pi /2\sqrt{2}r_{0} $ is determined from the boundary condition $ \delta \varphi_{\text{vac}}(r) = 0 $ at $ r \rightarrow \infty $.
   
It follows from this approximation that the first order RC $ \delta \varphi_{vac}(r) $ at $ r \gg r_{0} $ diminishes by the power law $ \sim 1/r^{5} $ (cf. Eqs. (\ref{PP}) and (\ref{fi}) where the behavior is exponential). 

Note also that values $ \bm{\mathcal{P}} $ and $ \bm{\mathcal{M}} $ become zero for electromagnetic plane waves.

\section{ Effective Hamiltonian of the Dirac atom}\label{Eff-Hamilt} 

In QED the DE is the Euler-Lagrange equation which follows from minimization of the action functional $ S = \int \mathcal{L} d^{4}x $ where $ \mathcal{L} $ is the Lagrange density which depends on the spinor and electromagnetic fields variables. The Dirac spinor field is described by bispinor $ \Psi $ and the Dirac conjugated one $ \bar{\Psi} = \Psi^{\dagger}\hat{\beta} $. The electromagnetic field variable is the 4-potential $ \vec{A} = \left( \varphi, \mathbf{A} \right) $ whose contravariant components $ A^{\mu} $ ($ \mu = 0,1,2,3 $) are the scalar $ \varphi $ and vector $ \mathbf{A} $ potentials as above.

The Hamilton functional of the spinor field is determined as
\begin{equation}
\label{H} 
\mathrm{H} = \int \left( i\hbar \Psi^{\dagger} \frac{\partial \Psi}{\partial t} - \mathcal{L} \right) d\mathbf{r} = \int \mathcal{H} d\mathbf{r} 
\end{equation} 
%= \int \Psi^{\dagger}(\mathbf{r}) \hat{H}_{D} \Psi (\mathbf{r}) d\mathbf{r} 
where $ \mathcal{H} = \Psi^{\dagger}(\mathbf{r}) \hat{H}_{D} \Psi (\mathbf{r}) $ is the Hamiltonian density with $ \hat{H}_{D} $ being the Dirac Hamiltonian, and the spatial integration is carried out over the whole volume. With account of the described above RC, the Dirac Hamiltonian in the presence of the Coulomb potential $ \varphi_{C} (\mathbf{r}) $ created by the external point charge $ e $ is 
\begin{equation}
\label{H_D} 
\hat{H}_{D} = c \bm{\hat{\alpha}} \cdot \hat{\mathbf{p}} + m c^{2} \hat{\beta} - \frac{e^{2}}{r} + V_{RC} \left(\mathbf{r} \right). 
\end{equation}
Here 
$ \bm{\hat{\alpha}} = \sum_{j} \mathbf{e}_{j} \hat{\alpha}_{j} $ is the vector matrix whose components $ \hat{\alpha}_{j} $ ($ j=x,y,z $) together with the matrix $ \hat{\beta} $ are the hermitian Dirac matrices  which in the standard representation have the following form
\begin{eqnarray}
\label{Dm}
\bm{\hat{\alpha}} = \left( 
\begin{array}{cc}
0 & \bm{\hat{\sigma}} \\ \bm{\hat{\sigma}} & 0
\end{array} \right),\qquad \:\hat{\beta} = \left( \begin{array}{cc}
\hat{I}_2 & 0 \\ 0 & -\hat{I}_2
\end{array} \right),
\end{eqnarray}
where $ \hat{I}_2 $ is a unit $ 2\times 2 $ matrix. Finally, the potential in the Dirac Hamiltonian (\ref{H_D}) is defined as
\[
V_{RC} \left(\mathbf{r}\right) = \]\[=-e\delta \varphi_{PP} (r) - e\delta \varphi_{elec} (r) - e\delta \varphi_{vac} (r) - e\delta\varphi_{mag} \left( \mathbf{r} \right) . 
\] 

The bispinor $ \Psi(\mathbf{r}) $ in Eq. (\ref{H}) is the amplitude of the Dirac spinor field and can be expanded over the complete ortho-normalized bispinor system. The  operator (\ref{H_D}) can be represented as $ \hat{H}_{D} = \hat{H}_{0} + V_{RC} $ where $ \hat{H}_{0} = c \bm{\hat{\alpha}} \cdot \mathbf{\hat{p}} + m c^{2} \hat{\beta} - e^{2}/r $ is the Dirac Hamiltonian with the Coulomb potential only. The DE with the Coulomb potential $ \hat{H}_{0} \Psi = E \Psi $  admits the exact solution and gives the complete ortho-normalized system of the eigen bispinors and the eigenvalue spectrum which includes the sectors of positive values $ E_{e,\lbrace \nu \rbrace} > 0 $ with the eigen bispinors $ \Psi_{e,\lbrace \nu \rbrace} $ (electrons) and of negative values $ E_{p,\lbrace \mu \rbrace} < 0 $ with the eigen bispinors $ \Psi_{p,\lbrace \mu \rbrace} $ (positrons) which  are characterized by their own sets of quantum numbers $ \lbrace \nu \rbrace $ and $ \lbrace \mu \rbrace $.

The expansion $ \Psi (\mathbf{r}) $ in Eq. (\ref{H}) over this ortho-normalized system of bispinors 
\begin{equation}
\label{Psiexpans} 
\Psi \left( \mathbf{r} \right) = \sum_{\lbrace \nu \rbrace} a_{\lbrace \nu \rbrace} \Psi_{e,\lbrace \nu \rbrace} \left( \mathbf{r} \right) + \sum_{\lbrace \mu \rbrace} b^{\dagger}_{\lbrace \mu \rbrace} \Psi_{p,\lbrace \mu \rbrace} \left( \mathbf{r} \right) ,
\end{equation} 
gives the relativistic Hamiltonian (\ref{H}) in the representation of the occupation numbers of the stationary states of the Hamiltonian $ \hat{H}_{0} $ with $ a^{\dagger}_{\lbrace \nu \rbrace} $ ($ a_{\lbrace \nu \rbrace} $) and $ b_{\lbrace \mu \rbrace}^{\dagger} $ ($ b_{\lbrace \mu \rbrace} $) being the creation (annihilation)  Fermi operators of a particle (an electron) and an antiparticle (a positron) in the state with quantum numbers $ \lbrace \nu \rbrace $ and $ \lbrace \mu \rbrace $, respectively. Substituting expression (\ref{Psiexpans}) into the functional (\ref{H}) and excluding the infinite  energy of the vacuum state, we come to the relativistic positively determined  operator 
\begin{equation}
	\label{Heff} 
\mathrm{H} = \mathrm{H}_{e} + \mathrm{H}_{p} + \hat{V}_{e-p}. 
\end{equation} 
Here $\mathrm{H}_{e}$, $\mathrm{H}_{p}$  and $ \hat{V}_{e-p} $ are the Hamiltonians of electrons  and positrons   in the Coulomb potential with the perturbation $ V_{RC} $, and the operator of their mutual transformations due to the perturbation, respectively.

 For any non-relativistic perturbation, the renormalization of electron and positron states is negligibly small which allows to consider particles and antiparticles as non-interacting objects. 
For a positive point charge $ e $ the negative eigenvalues of $ \hat{H}_{0} $ include the continuous spectrum only, and positron eigenstates are the spherical waves. Positive eigenvalues of the DE determine the electron energies which include continuous spectrum of unbound electrons and discrete electron levels. Below we consider the discrete spectrum, only. It is characterized by the quantum numbers
\begin{equation}
	\label{QN}
	\lbrace \nu \rbrace = n,j,m_{j},\sigma  
\end{equation} 
in which 
\begin{eqnarray}
	\label{QN1}
	n = 1,2,\ldots , \quad j = \frac{1}{2},\, \frac{3}{2},\ldots\:,n-\frac{1}{2},\\
	m_{j} = \pm j,\pm (j-1),\ldots ,\pm \frac{1}{2} \nonumber 
\end{eqnarray}
are the principal quantum number, the total angular momentum and its projection on the polar axis, respectively, and $ \sigma $ is the sign of the spinor invariant (see \cite{BEL-gen}). 

The eigen bispinors of the stationary bound electron states in an atom are the well-known Darwin solutions which can be written down as 
\begin{equation}
\label{bisp_sigma} 
\Psi_{\lbrace \nu \rbrace  } (\mathbf{r}) = {\begin{pmatrix}R_{n,j}^{(\sigma)} (r) \Omega_{j-\sigma (1/2), m_{j},\sigma}(\vartheta ,\varphi)\\ i \lambda_{n,j} Q_{n,j}^{(-\sigma)}(r) \hat{\sigma}_{r} \Omega_{j-\sigma (1/2), m_{j},\sigma}(\vartheta ,\varphi)\\\end{pmatrix}}
\end{equation} 
where $ \Omega_{j\mp1/2, m_{j},\pm} $ are the spherical spinors, functions $ R_{n_{r},j}^{(\pm)} $ and $ G_{n_{r},j}^{(\mp)} $ are the radial functions whose explicit expressions are given in \textbf{Appendix}, and parameter $ \lambda_{n,j} \ll 1 $ indicates the smallness of the lower spinor in the bispinor $ \Psi_{n, j, m_{j},\sigma } $ (see below). According to the complete set of the integrals of motion, these bispinors are characterized by the quantum numbers (\ref{QN}), (\ref{QN1}), where  $ \sigma = \pm $ indicates the states with positive or negative (equal to $  \sigma \kappa_{j} \equiv \sigma (j+1/2) $) eigenvalue of the Dirac invariant. 

Hydrogen atom levels are  given by the expression
\begin{equation}
\label{E_n,j}
\varepsilon_{n,j} = \sqrt{1 - \frac{\alpha^{2}}{\mathcal{N}_{n,j}^{2}}} = \frac{n - \Delta_{j}}{\mathcal{N}_{n,j}}, \end{equation}
where 
\begin{equation}
	\label{N}
 \mathcal{N}_{n,j} = \sqrt{\left(n - \Delta_{j} \right)^{2} +  \alpha^{2}}, 
\end{equation}
\begin{equation}
\label{delta_n,j} 
\Delta_{j} = \kappa_{j} - \gamma_{\kappa} = \frac{\alpha^{2}}{\kappa_{j} + \gamma_{\kappa}},
\end{equation}
\begin{equation}
	\label{gamma} 
	\gamma_{\kappa} = \sqrt{\kappa_{j}^{2} -\alpha^{2}}, \quad \kappa_{j} = j + \frac{1}{2}. 
\end{equation}
Here $ \Delta_{j} $ is the splitting of  the fine-structure multiplets which is usually treated  as arising due to the spin-orbit interaction. 
Parameter $ \lambda_{n,j} $ 
\begin{equation}
	\label{lambda_j} 
	\lambda_{n,j} = \sqrt{ \frac{1 - \varepsilon_{n,j}}{1 + \varepsilon_{n,j}} } = \frac{\alpha}{\mathcal{N}_{n,j} + n - \Delta_{j} }.
\end{equation} 
in the bispinor (\ref{bisp_sigma}) determines the relation between the upper and lower spinors of the Dirac bispinor.

Therefore, the Hamiltonian of the  Dirac atom bound states is 
\[
\mathrm{H} = \sum_{n,j,m_{j},\sigma} E_{n,j} a_{n,j,m_{j},\sigma}^{\dagger} a_{n,j,m_{j},\sigma} +
\]
\begin{equation}
	\label{H_DA} 
\quad + \sum_{\lbrace \nu\rbrace,\lbrace \nu'\rbrace} V_{\lbrace \nu\rbrace,\lbrace \nu'\rbrace} a_{\lbrace \nu\rbrace}^{\dagger} a_{\lbrace \nu'\rbrace},
\end{equation}
where $ E_{n,j} = mc^{2} \varepsilon_{n,j} $ and  
\begin{equation}
\label{matrixel} 
V_{\lbrace \nu \rbrace ,\lbrace \nu' \rbrace } =  \int \Psi_{\lbrace \nu \rbrace }^{\dagger}(\mathbf{r}) V_{RC} \left( \mathbf{r} \right) \Psi_{\lbrace \nu' \rbrace } (\mathbf{r}) d^{3}r
\end{equation} 
are matrix elements of the RC.

\section{Matrix elements of the radiative corrections}\label{matr-el}

The RC in Eq. (\ref{H_D}) can be represented as the sum of two terms, 
\[V_{RC} ( \mathbf{r}) = V^{(1)} ( \mathbf{r}) + V^{(2)} ( \mathbf{r})  \]
where 
\[ V^{(1)} ( \mathbf{r}) \equiv V^{(1)}(r)= -e\delta \varphi_{PP} (r) - e\delta \varphi_{elec} (r) - e\delta \varphi_{vac} (r)\equiv  \]
\[\quad \equiv V_{PP}( r)+V_{elec}( r)+V_{vac}( r) = V^{(1)} ( r) \]
(see Eqs. (\ref{Coul_pert}), (\ref{deltafi_elec}) and (\ref{deltafi_vac})), and 
\[ V^{(2)} ( \mathbf{r})= - \frac{ e^{2} \alpha \lambda_{C}}{4\pi } \frac{d\phi \left(r \right)}{dr} \bm{\hat{\Gamma}} \cdot \mathbf{e}_{\mathbf{r}} \equiv V_{mag}( \mathbf{r}) = V^{(2)} ( r) \bm{\hat{\Gamma}} \cdot \mathbf{e}_{\mathbf{r}} \]
includes the correction from the magnetic electron form factor (\ref{deltafi_magn}).

The perturbation $ V^{(1)} ( \mathbf{r}) =  V^{(1)} (r) $ is the scalar potential whose matrix elements are 
\[
\begin{array}{c} 
	V_{\lbrace \nu \rbrace ,\lbrace \nu' \rbrace }^{(1)} = 
	\int V^{(1)} ( r) \Psi_{\lbrace \nu \rbrace  }^{\dagger}(\mathbf{r}) \Psi_{\lbrace \nu' \rbrace }(\mathbf{r}) d^{3}r = \\ 
	=\int d\varphi \sin \vartheta d\vartheta r^{2} dr  V^{(1)} (r)  \left[ R_{n,j}^{(\sigma)}(\rho) R_{n',j'}^{(\sigma')}(\rho')\right. + \\
	\left. + \lambda_{n,j} \lambda_{n',j'} Q_{n_,j}^{(-\sigma)} (\rho) Q_{n',j'}^{(-\sigma')} (\rho') \right] \times \\ 
	\qquad  \times \Omega_{j-\sigma 1/2, m_{j},\sigma}^{\dagger}\left(\vartheta ,\varphi \right) \Omega_{j'-\sigma' 1/2, m'_{j},\sigma'} \left(\vartheta ,\varphi \right) . 
\end{array} 
\]
Since the spherical spinors are orthogonal, the matrix elements in the Hamiltonian (\ref{H_DA}) are diagonal: $ V_{\lbrace \nu \rbrace ,\lbrace \nu' \rbrace }^{(1)}  = V_{j,\sigma ;n,n'}^{(1)} \delta_{j',j} \delta_{m'_{j},m_{j}} \delta_{\sigma',\sigma} $. 

According to (\ref{deltafi_magn}),  matrix elements of the perturbation $ V_2 ( \mathbf{r})$ are 
\[
V^{(2)}_{\lbrace \nu \rbrace ,\lbrace \nu' \rbrace } = 
\int \Psi_{\lbrace \nu \rbrace }^{\dagger}(\mathbf{r}) \Phi (r) \bm{\hat{\Gamma}} \cdot \mathbf{e}_{\mathbf{r}} \Psi_{\lbrace \nu' \rbrace }(\mathbf{r}) d^{3}r = 
\] 
\[ = \int V^{(2)} (r) 
\left[ \lambda_{n',j'} R_{n,j}^{(\sigma)}(\rho) Q_{n',j'}^{(-\sigma')}(\rho') + \lambda_{n,j}  Q_{n_,j}^{(-\sigma)} (\rho) R_{n',j'}^{(\sigma')} (\rho') \right] \times 
\] 
\[  \times \Omega_{j-\sigma 1/2, m_{j},\sigma}^{\dagger}\left(\vartheta ,\varphi \right)  \Omega_{j'-\sigma' 1/2, m'_{j},\sigma'} \left(\vartheta ,\varphi \right)  d^3r=
\]
\[ \quad = V^{(2)}_{j,\sigma ;n,n'}\delta_{j',j} \delta_{m'_{j},m_{j}}  \delta_{\sigma',\sigma}. 
\]
and also are diagonal according numbers $ j,m_{j},\sigma $. 

Therefore, the Hamiltonian of the bound hydrogen states with account of the RC is 
\[
\mathrm{H} = \sum_{n,j,m_{j},\sigma} \left( E_{n,j} + V_{n,j,\sigma} \right) a_{n,j,m_{j},\sigma}^{\dagger} a_{n,j,m_{j},\sigma} + 
\]
\[ \qquad + \sum_{n\neq n',j,m_{j},\sigma} V_{n,n';j,\sigma} a_{n,j,m_{j},\sigma}^{\dagger} a_{n',j,m_{j},\sigma}.
\] 
As any quadratic form, this Hamiltonian can be exactly diagonalized. But the main input into the energy shift is due to the diagonal elements of $ V^{(1)} ( \mathbf{r})$ and $V^{(2)} ( \mathbf{r})$ which will be calculated below.

With account of the explicit expressions for the radial functions (\ref{radFunc})-(\ref{R_n,j}),
the diagonal elements $ V^{(i)}_{n,j,\sigma } $ ($ i = 1,2 $) are  
\[
V^{(i)}_{n,j,\sigma} = A_{n,j}^{2} \int_{0}^{\infty} V_{RC}^{(i)}  \left( \rho \right) e^{-\rho } \rho^{2\gamma_{j}} K_{n_{r},j,\sigma}^{(i)}(\rho ) d\rho , \quad i = 1,2
\]
where $ K_{n_{r},j,\sigma}^{(i)} $ are polynomials of the order $ 2n_{r} $: 
\begin{equation}
	\label{K^1_n,j}  
	K_{n_{r},j,\sigma}^{(1)}(\rho ) = \left( P_{n_{r},j}^{(\sigma)} (\rho )\right) ^2  + \lambda_{n,j}^{2}  \left( W_{n_{r},j}^{(-\sigma)}(\rho )\right)^2 = 
	\sum_{\nu = 0}^{2n_{r}} a_{\nu}^{(1)}(n,j,\sigma) \rho^{\nu} , 	
\end{equation} 
and 
\begin{equation}
	\label{K^2_n} 	
	K_{n_{r},j,\sigma}^{(2)}(\rho ) = 	P_{n_{r},j}^{(\sigma)} (\rho ) W_{n_{r},j}^{(-\sigma)}(\rho )  = 
	\sum_{\nu = 0}^{2n_{r}} a_{\nu}^{(2)}(n,j,\sigma) \rho^{\nu} . 	
\end{equation}
Here coefficients $ a_{\nu}^{(i)}(n,j,\sigma) $ depend on the explicit expressions of the radial functions for given state. Therefore the diagonal elements are  
\begin{equation}
	\label{V^i_n} 
	V^{(i)}_{n,j,\sigma} = A_{n,j}^{2} \sum_{\nu = 0}^{2n_{r}} a_{\nu}^{(i)}(n,j,\sigma)
	\int_{0}^{\infty} V_{RC}^{(i)}  \left( \rho \right) e^{-\rho } \rho^{2\gamma_{j} + \nu} d\rho , \quad i = 1,2 \: . 
\end{equation}

\vspace{5mm}

\subsection{\bf{Modification of the Coulomb field by the photon polarization}}\label{m-e-PP} 

In this case taking into account Eq. (\ref{Coul_pert})  the PP perturbation 
can be written as 
\begin{equation}
\label{V_U} 
-e\delta \varphi_{PP} (\rho)  = -\frac{4mc^{2}\alpha^{3}}{3\pi \mathcal{N}_{n,j}} \frac{1}{\rho} F (\rho)
\end{equation}
where 
\[
 F (\rho) = \int_{1}^{\infty} e^{-( \mathcal{N}_{n,j} /\alpha) \rho \zeta} \left( 1 + \frac{1}{2\zeta^{2}} \right) \frac{\sqrt{\zeta^{2} - 1}}{\zeta^{2}} d\zeta
 \]
and
\begin{equation}
	\label{rho}
	\rho = \frac{2r}{r_{B} \mathcal{N}_{n_{r},j}} 
\end{equation} 
is the dimensionless radial variable, characteristic for each energy state. Here $ r_{B} = \hbar^{2} /me^{2} $ is the Bohr radius.

The diagonal matrix elements (\ref{V^i_n}) $ V^{(1)}_{n,j,\sigma} = V_{PP}\left( n,j,\sigma \right) $ after integration over $ \rho $ are given by the expressions
\[
V_{PP}\left( n,j,\sigma \right) = -\frac{mc^{2}\alpha^{3}}{3\pi }
\frac{\left( 1 + \varepsilon_{n,j} \right) \left( \mathcal{N}_{n,j} + \kappa_{j} \right) n_{r}! }{\mathcal{N}_{n,j}^{2} \Gamma \left( n_{r} + 1 + 2\gamma_{j} \right)} 	\times 
\]
\begin{equation}
	\label{V^U_n,j} 
	\times  \sum_{\nu = 0}^{2n_{r}} a_{\nu}^{(1)}(n,j,\sigma) \Gamma \left(2\gamma_{j} + \nu \right) \left(\frac{\alpha}{\mathcal{N}_{n,j}} \right)^{2\gamma_{j} + \nu} C_{2\gamma_{j} + \nu}^{(PP)}, 
\end{equation} 
where 
\begin{equation}
	\label{cmu}
	C_{\mu}^{(PP)} = \int_{1}^{\infty} \left( 1 + \frac{1}{2\zeta^{2}} \right) \frac{\sqrt{\zeta^{2} - 1}}{\zeta^{2}\left(\zeta + \alpha /\mathcal{N}_{n,j} \right)^{\mu} } d\zeta . 
\end{equation}

Due to smallness of $ \alpha \ll 1 $ , it is possible to use the approximation of $ \gamma_{j} $  by integral number $ \gamma_{j} \simeq \kappa_{j} $ and in view of the inequality $ \zeta \gg \alpha /\mathcal{N}_{n,j} $ ($ \zeta \geq 1 $) the series expansion 
\[
\left( \zeta + \alpha /\mathcal{N}_{n,j} \right)^{-\mu} \approx \zeta^{-\mu} \left[ 1 - \mu \frac{\alpha}{\mathcal{N}_{n,j}} + \frac{\mu (\mu + 1)}{2!} \left( \frac{\alpha}{\mathcal{N}_{n,j}} \right)^{2} \frac{1}{\zeta^{2}} - \ldots \right] 
\]
can be used under the integral in Eq. (\ref{cmu}). 

The explicit expressions of the Uehling potential diagonal matrix elements for the lowest energy states are given below.

i) For the ground state $ 1S_{1/2} $ with quantum numbers $ n = 1, \: j=1/2, \: \sigma = + $  ($ n_{r} = 0 $, $ \kappa_{1/2} = 1 $) we have following values
\[
\mathcal{N}_{1,1/2} = 1 , \quad \mu = 2\gamma_{1} \simeq 2 , \quad a_{0}^{(1)}(1,1/2,+) = \frac{2}{1 + \varepsilon_{1,1/2}} ,
\]
which give $ C_{2}^{(PP)} \cong 2/5 $ and 
\[
V_{PP}\left( 1S_{1/2} \right)  \approx -\frac{4mc^{2}\alpha^{3+2\gamma_{1}}}{15\pi \gamma_{1}} \simeq  -\frac{4mc^{2}\alpha^{5}}{15\pi } .
\]

ii) The excited state $ 2P_{3/2} $ with quantum numbers $ n = 2, \: j = 3/2, \: \sigma = +  $ ($ n_{r} = 0 $, $ \kappa_{3/2} = 2 $) is characterized by numbers 
\[
\mathcal{N}_{2,3/2} = 2 , \quad \mu = 2\gamma_{2} \simeq 4 , \quad a_{0}^{(1)}(2,3/2,+) = \frac{2}{1 + \varepsilon_{2,3/2}} ,
\]
which in the same approximation gives $ C_{4}^{(PP)} \simeq 6/35 $ and   
\[
V_{PP}\left( 2P_{3/2} \right) \approx - \frac{mc^{2}\alpha^{7}}{560\pi}  .
\] 

iii) For the state $ 2S_{1/2} $  with quantum numbers $ n = 2, \: j = 1/2, \: \sigma = +  $ ($ n_{r} = 1 $, $ \kappa_{1/2} = 1 $) and the radial function polynomial (\ref{K^1_n,j}) 
\begin{equation}
	\label{K(2S_1/2)} 
	K_{1,1/2,+}^{(1)} = \frac{2}{1 + \varepsilon_{2,1/2}} \left[ \left( \mathcal{N}_{2,1/2} - 1 \right)^{2} \left( \mathcal{N}_{2,1/2} + 2 \right) - \left( \mathcal{N}_{2,1/2} - 1 \right) \left( \mathcal{N}_{2,1/2} + 2 \right) \rho + \rho^{2} \right] . 
\end{equation}
the matrix element in Eq. (\ref{V^U_n,j}) in the approximation $ \mathcal{N}_{2,1/2} \simeq 2 $ and $ \gamma_{1} \simeq 1 $ is 
\[
V_{PP}\left( 2S_{1/2} \right) \simeq -\frac{mc^{2}\alpha^{5} }{6\pi } \left( \frac{1}{5} - \frac{15\pi}{128} \alpha  \right)  
\]

iv) For the state $ 2P_{1/2} $  with quantum numbers $ n = 2, \: j = 1/2, \: \sigma = -  $ ($ n_{r} = 1 $, $ \kappa_{1/2} = 1 $) the radial function polynomial (\ref{K^1_n,j}) is 
\[
K_{1,1/2,-}^{1}(\rho ) = \frac{2\left(1+2\gamma_{1} \right) }{1 + \varepsilon_{2,1/2}} \left[ \left( 2 - \mathcal{N}_{2,1/2} \right) + \frac{2 - \mathcal{N}_{2,1/2}}{ \mathcal{N}_{2,1/2} + 1} \rho + \frac{\rho^{2}}{\left(\mathcal{N}_{2,1/2} + 1 \right)^{2}} \right] . 
\]
Here $ \mathcal{N}_{2,1/2} \approx 2 $ and $ 2 - \mathcal{N}_{2,1/2} \simeq \alpha^{2}/4 $. Therefore, the polynomial can be approximated by the expression  
\begin{equation}
	\label{K(2P_1/2)} 
	K_{1,1/2,-}^{1}(\rho ) \simeq \frac{2\left(1+2\gamma_{1} \right) }{1 + \varepsilon_{2,1/2}} 
	\left[ \frac{\alpha^{2}}{4} 
	+ \frac{\alpha^{2}}{12} \rho 
	+ \frac{1}{9} \rho^{2} \right]  
\end{equation}
and for the matrix element (\ref{V^U_n,j}) one can obtain the expression
\[
V_{PP}\left( 2,1/2,- \right) =-\frac{mc^{2}\alpha^{7}}{32 \pi } \left( \frac{9}{35} +  \frac{5\pi}{128} \alpha\right) . 
\] 

\vspace{5mm}

\subsection{\bf{Modification of the Coulomb field by the electron electric form factor}}\label{EFF}

The expression for the electric form factor $ F_{1} (q^{2} ) $ in Eq. (\ref{vert-corr}) contains in the argument of the logarithm the parameter $ \lambda \rightarrow 0 $ which is "finite virtual photon mass" introduced to remove the infrared singularity. It plays the role of a low-frequency cutoff parameter. In the standard calculation, this parameter in the general case of arbitrary $Z$  is assumed to be in the interval $ \left( Z\alpha \right)^{2}m \ll \lambda \ll m $ and after addition of the low-frequency contribution parameter  $ \lambda $ cancels out. Minimization of the low-frequency contribution leads to the equality $ \ln \left( 4m^{2} /\lambda^{2} \right) = 4 \ln \left( 1/\alpha \right) + const $, where  $const$ is close to 1 \cite{RelQuant,Akhiezer,Flambaum} for $Z\alpha \ll 1$. 

The SE effective potentials (\ref{deltafi_elec}) and (\ref{deltafi_magn}) were first considered in \cite{Flambaum} and used in calculations of the  relativistic heavy elements and in molecular calculations (see \cite{Ginges,Sunaga}). As above, here we consider the hydrogen spectrum ($Z=1$), so  
\begin{equation}
	\label{lambda_ef} 
\ln \frac{2m}{\lambda} = 2 \ln {\frac{1}{\alpha}} . 
\end{equation} 

According to Eq. (\ref{deltafi_elec}), the perturbation $ V_{elec} (\rho) = -e\delta \varphi_{E} (\rho) $ acquires the form
\begin{equation}
	\label{V_E} 
	V_{elec} (\rho) = \frac{mc^{2}\alpha^{3}}{\pi \mathcal{N}_{n,j}} \frac{1}{\rho} F_{2} (\rho) ,  
\end{equation}  
where 
\[
F_{2} (\rho) = \int_{1}^{\infty} d\zeta f_{elec}(\zeta) e^{-\frac{\mathcal{N}_{n,j}}{\alpha} \zeta\rho},
\] 
\[f_{elec}(\zeta) = \frac{2 - 3\zeta^{2} + \left(2\zeta^{2} - 1\right) \ln [\left(\frac{4m}{\lambda} \right)^{2} \left(\zeta^{2} -1\right)] }{\zeta^{2}\sqrt{\zeta^{2} - 1}}. 
\]  
So, the diagonal matrix elements determined in Eq. (\ref{V_E}), are 
\[
V_{elec}\left( n,j,\sigma \right) = -\frac{mc^{2}\alpha^{3}}{4\pi }
\frac{\left( 1 + \varepsilon_{n,j} \right) \left( \mathcal{N}_{n,j} + \kappa_{j} \right) n_{r}! }{\mathcal{N}_{n,j}^{2} \Gamma \left( n_{r} + 1 + 2\gamma_{j} \right)} 	\times 
\]
\begin{equation}
	\label{V_elec} 
	\times  \sum_{\nu = 0}^{2n_{r}} a_{\nu}^{(1)}(n,j,\sigma) \Gamma \left(2\gamma_{j} + \nu \right) \left(\frac{\alpha}{\mathcal{N}_{n,j}} \right)^{2\gamma_{j} + \nu} C^{(elec)}_{2\gamma_{j} + \nu}, 
\end{equation} 
where 
\[
C^{(elec)}_{\mu} = \int_{1}^{\infty} \frac{f_{elec}(\zeta)}{\left(\zeta + \alpha /\mathcal{N}_{n,j} \right)^{\mu} } d\zeta . \]
Again in these integrals we can use the series expansion of the expression 
$ \left( \zeta + \alpha /\mathcal{N}_{n,j} \right)^{-\mu} $ and calculate the integrals using the same change of the variable as in the case of the electron form-factors 
 \cite{RelQuant}. 

For states with the maximal possible value $ j = j_{n} = n - 1/2 $ at the given $ n $ when $ K_{n}^{(+)} = 2/(1+\varepsilon_{n}) $  we have 
\[
V_{elec}(n,+) = 
\frac{mc^{2}\alpha^{2\gamma_{n}+3}}{2\pi n^{2\gamma_{n}+1}\gamma_{n}}  
C^{(elec)}_{2\gamma_{n}} , \quad \gamma_{n} \simeq n .
\] 

Below we write down the explicit expressions of the diagonal matrix elements for the lowest energy states in the approximation $ \mathcal{N}_{2,1/2} \simeq 2 $ and $ \gamma_{1} \simeq 1 $.

i)  For the ground state $ 1S_{1/2} $ accounting for Eq.  (\ref{lambda_ef}), one has 
\[
C^{(elec)}_{2} \simeq  \frac{8}{3} \left( 2\ln \frac{1}{\alpha} - \frac{3}{8} \right).
\] 
Therefore,  
\[
 V_{elec}(1S_{1/2})= 
\frac{4mc^{2}\alpha^{5}}{3\pi } \left( 2 \ln \frac{1}{\alpha} - \frac{3}{8} \right).
\] 

ii) For the first excited state  $ 2P_{3/2} $ (i.e., $ n = 2,\: j = 3/2,\: \sigma = + $) one has
\[
C^{(elec)}_{4} = 
%\frac{2}{5} \int_{1}^{\infty} \frac{d\zeta}{\zeta^{4}\sqrt{\zeta^{2}-1}} + \]
%\[\qquad + \frac{1}{5} \int_{1}^{\infty} \frac{-7 + 6 \ln [\left(\frac{4m}{\lambda} %\right)^{2}(\zeta^{2}-1)]}{\zeta^{4}\sqrt{\zeta^{2}-1}} =\]
%\[\qquad  =
\frac{8}{5} \left( 2 \ln \frac{1}{\alpha} - \frac{11}{12} \right) 
\] 
and, therefore, 
\[
V_{elec}(2P_{3/2}) = \frac{mc^{2}\alpha^{7}}{80 \pi } 
\left( 2 \ln \frac{1}{\alpha} - \frac{11}{12} \right)   .
\] 

iii) For the state $ 2S_{1/2} $ ($ n = 2, \: j=1/2 ,\: \sigma = + $ and $ n_{r}=1 $) with radial function polynomial (\ref{K(2S_1/2)}) and 
\[
C^{(elec)}_{3} \approx \frac{\pi}{8} \left( 10 \ln \frac{1}{\alpha} - 7 \right) 
\] 
it can be found that the diagonal matrix element is
\[
V_{elec}(2S_{1/2}) = 
\frac{mc^{2}\alpha^{5}}{6 \pi }   
\left[ 2\ln \frac{1}{\alpha} - \frac{3}{8}  
- \frac{3\pi}{64} \left( 10 \ln \frac{1}{\alpha} - 7 \right) \alpha  \right]  
\]

iv) For the state $ 2P_{1/2} $ ($ n = 2, \: j=1/2 ,\: \sigma = - $ and $ n_{r}=1 $) with $ K_{1,1/2,-}^{(1)}(\rho ) $ (\ref{K(2P_1/2)}) one gets the corresponding diagonal matrix element for the electric form factor
\[
V_{elec}(2P_{1/2}) \simeq  \frac{mc^{2}\alpha^{7}}{2^{4}\pi }      
\left( \frac{7}{5} \ln \frac{1}{\alpha} 
- \frac{89}{240}  + \frac{\alpha }{6} C^{(elec)}_{3} \right).
\]

\vspace{5mm}

\subsection{\bf{Modification of the Coulomb field by the vacuum polarization due to the nucleus charge field}}\label{vac-pol}

This perturbation is described by the term $ V_{vac}(r) = -e\delta  \varphi_{vac}(r) $ which, taking into account  expression (\ref{deltafi_vac}), reads as
\[
V_{vac}(\rho) = \frac{2mc^{2}\alpha^{2}}{3 \mathcal{N}_{n,j}} \tilde{\Phi} (\rho).
\] 
Here, according to (\ref{F(r)}), 
\[
\tilde{\Phi} (\rho) = \frac{1}{\rho} - \frac{\pi}{2\sqrt{2}\beta_{n,j}} - 
\]
\begin{equation}
	\label{tildePhi} 
	- \frac{1}{4\sqrt{2} \beta_{n,j}} \frac{1}{2}\ln \frac{\rho^{2} + \sqrt{2} \beta_{n,j} \rho + \beta_{n,j}^{2}}{\rho^{2} - \sqrt{2} \beta_{n,j} \rho + \beta_{n,j}^{2}} + 
\end{equation}
\[
+ \frac{1}{2\sqrt{2} \beta_{n,j}} \left[ \arctan \left( \frac{\sqrt{2}\rho}{\beta_{n,j}} - 1 \right) + \arctan \left( \frac{\sqrt{2}\rho}{\beta_{n,j}} + 1 \right) \right],
\]
and the following notation is introduced
\[
\beta_{n,j} = \frac{2\alpha}{\mathcal{N}_{n,j}} \left( \frac{2 \alpha^{3}}{15 \pi } \right)^{1/4}.   
\]

The diagonal matrix elements of this perturbation are given by the expressions
\begin{equation}
	\label{DME-vac}
	V_{vac}(n,j,\sigma) = \frac{2mc^{2}\alpha^{2}}{3\mathcal{N}_{n,j} }
	A_{n,j}^{2} \mathcal{I}_{n,j}^{(\sigma)}  
\end{equation}
where 
\[
\mathcal{I}_{n,j}^{(\sigma)} = \int_{0}^{\infty} \tilde{\Phi} (\rho) e^{-\rho } \rho^{2\gamma_{j}} K_{n_{r},j,\sigma}^{(1)}(\rho ) d\rho 
\]
where $ \tilde{\Phi} $ is defined  in Eq. (\ref{tildePhi}) and $ K_{n_{r},j,\sigma}^{(1)} $ is a polynomial (\ref{K^1_n,j}). Substituting the explicit expressions of these polynomials into the matrix element, they take the form  
\[
\mathcal{I}_{n,j}^{(\sigma)} 
= \sum_{\nu = 0}^{2n_{r}} a_{\nu}^{(1)}(n,j,\sigma) \int_{0}^{\infty} \tilde{\Phi} (\rho) e^{-\rho } \rho^{2\gamma_{j}+\nu} d\rho .
\] 
Therefore, the problem is reduced to calculation of integrals  
\begin{equation}
	\label{calJ_mu} 
	\mathcal{J}_{2\kappa_{j}+\nu} = \int_{0}^{\infty} \tilde{\Phi} (\rho)  e^{-\rho } \rho^{2\gamma_{j}+\nu} d\rho 
\end{equation}
where $ \tilde{\Phi} $ is given in Eq. (\ref{tildePhi}). The degree $ 2\gamma_{j}+\nu = \mu $ of the variable $ \rho $ can be approximated by integer numbers $ \mu = 2\kappa_{j}+\nu $ in view of the smallness of the constant $ \alpha \ll 1 $ ($ \gamma_{j} \simeq \kappa_{j} $). But even in this case we do not have analytical expression for such integrals. Therefore, to obtain  the dependence of integrals $ \mathcal{J}_{2\kappa_{j}+\nu} $  on the small parameter $ \beta_{n,j} $  analytically, we present the logarithm and arctangent functions in Eq. (\ref{tildePhi}) via integrals:
\[
\frac{1}{4\sqrt{2} \beta_{n,j}} \frac{1}{2}\ln \frac{\rho^{2} + \sqrt{2} \beta_{n,j} \rho + \beta_{n,j}^{2}}{\rho^{2} - \sqrt{2} \beta_{n,j} \rho + \beta_{n,j}^{2}} = \frac{1}{4} \int_{-1}^{1} dt \frac{ \rho}{\rho^{2} + \sqrt{2} \beta t \rho + \beta^{2} } ,
\]
and 
\[
\frac{1}{2\sqrt{2} \beta_{n,j}} \left[ \arctan \left( \frac{\sqrt{2}\rho}{\beta_{n,j}} - 1 \right) + \arctan \left( \frac{\sqrt{2}\rho}{\beta_{n,j}} + 1 \right) \right] = 
\] 
\[
= \frac{1}{4} \int_{0}^{1} dt \left[ 
\frac{\rho - \beta /\sqrt{2}}{ \beta^{2}/2 + \left( \rho - \beta /\sqrt{2} \right)^{2}t^{2} } + \frac{\rho + \beta /\sqrt{2}}{ \beta^{2} + \left( \rho + \beta /\sqrt{2} \right)^{2}t^{2} } \right] .
\]
Changing variable $ \rho $ by $ x = \rho \mp \beta/\sqrt{2} $,  one can perform integration over the  radial variable in eq. (\ref{calJ_mu}) which results in representation of the matrix elements of vacuum polarization (\ref{DME-vac}) via the following integrals 
\[
\int_{0}^{\infty}  \frac{e^{-x} x^{2n+1} } {x^{2} + a^{2} } dx = \left( -1 \right)^{n-1} a^{2n} \left[ \mathrm{ci} \left( a\right) \cos a + \mathrm{si} \left( a \right) \sin a \right]  
+ \sum_{k=1}^{n} (2n - 2k + 1)! \left( -a^{2} \right)^{k-1} 
\] 
if $ \mu = 2\kappa_{j}+\nu $ is odd number, and 
\[
\int_{0}^{\infty}  \frac{e^{-x} x^{2n} } {x^{2} + a^{2} } dx = \left( -1 \right)^{n} a^{2n-1} \left[ \mathrm{ci} \left( a\right) \sin a - \mathrm{si} \left( a \right) \cos a \right] 
+ \sum_{k=1}^{n} (2n - 2k )! \left( -a^{2} \right)^{k-1} 
\]
at even $ \mu $. Here $ \mathrm{si} (a) $ and $ \mathrm{ci} (a) $ are integral sine and  cosine  functions, respectively, and $ a \sim \beta_{n,j} $. Because $ \beta_{n,j} \ll 1 $, one can use for $ \mathrm{ci} (a) $ and $ \mathrm{si} (a) $ the expansions 
\[ 
\mathrm{si} (\xi) = - \frac{\pi}{2} + \sum_{k=1}^{\infty} \frac{(-1)^{k+1}\xi^{2k-1}}{\left( 2k-1 \right) \left( 2k-1 \right)! } , 
\]  
\[
\mathrm{ci} (\xi) = \mathbb{C} - \ln \xi + \sum_{k=1}^{\infty} \left( -1 \right)^{k} \frac{\xi^{2k}}{2k (2k)!} 
\]
where $ \mathbb{C} = 0.5772... $ is the Euler's constant. 

Below, omitting the cumbersome calculation details with the use of table integrals  \cite{Ryzhik}, we write down the explicit expressions of the diagonal matrix elements for the lowest energy states.

i) For the ground state $ 1S_{1/2} $ with the quantum numbers $ n = 1, \: j=1/2, \: \sigma = + $  ($ n_{r} = 0 $, $ \kappa_{1/2} = 1 $, $ \gamma_{1/2} \simeq 1 $), and  
\[
\mathcal{N}_{1,1/2} = 1 , \quad a^{1}_{0} = \frac{2}{1+\varepsilon_{1}} , \quad 
\beta_{1,1/2} \equiv \beta_{1} = 2\alpha \left( \frac{2 \alpha^{3}}{15 \pi } \right)^{1/4},
\]
the diagonal matrix element is 
\[
V_{vac}(1,1/2,+) = \frac{mc^{2}\alpha^{2}}{3 } \left( 
\frac{2}{3} \mathbb{C}  
+ \frac{2}{3}  \ln \frac{\sqrt{2}}{\beta_{1}} 
+ \frac{1}{3}\sqrt{2} \ln \frac{\sqrt{2} + 1}{\sqrt{2} - 1} 
+ \frac{5}{12} \ln 2 
+ \frac{\pi}{12} 
- \frac{1}{3} 
\right) \frac{\beta_{1}^{2}}{2}  .  
\]

ii) For the excited state $ 2P_{3/2} $ with $ n = 2,\: j = 3/2, \: \sigma = + $ ($ n_{r} = 0 $, $ \kappa_{3/2} = 2 $, $ \gamma_{3/2} \simeq 2 $) and   
\[
\mathcal{N}_{2,3/2} = 2 , \quad a^{(1)}_{0}(2,1/2) = \frac{2}{1+\varepsilon_{2}} , \quad 
\beta_{2,3/2} \equiv \beta_{2} = \alpha \left( \frac{2 \alpha^{3}}{15 \pi } \right)^{1/4}
\]
in the same approximation, the diagonal matrix element is
\[
V_{vac}(2,3/2,+) = - \frac{5mc^{2}\alpha^{2}}{216 }\left( 2 - \sqrt{2} \right) \frac{\beta_{2}^{2}}{2}  ,
\]  

iii) For the excited state $ 2S_{1/2} $ with $ n = 2,\: j = 1/2, \: \sigma = + $ ($ n_{r} = 1 $, $ \kappa_{j} = 1 $) in the approximation $ \gamma_{1/2} \simeq 1 $, and   
\[
\mathcal{N}_{2,1/2} \approx 2 , \quad K_{1,1/2}^{+}(\rho ) \simeq \frac{2}{1 + \varepsilon_{2,1/2}} \left( 4 - 4 \rho + \rho^{2} \right) , \quad \beta_{2,1/2} \approx \beta_{2} ,
\]
the diagonal matrix element is 
\[
V_{vac}(2,1/2,+) = \frac{mc^{2}\alpha^{2}}{24 } \left( 
\frac{8}{3} \mathbb{C}  
+ \frac{8}{3} \ln \frac{\sqrt{2}}{\beta_{2}} 
+ \frac{4}{3}\sqrt{2} \ln \frac{\sqrt{2} + 1}{\sqrt{2} - 1} 
+ \frac{5}{3} \ln 2 
+ \frac{\pi}{3} 
+ \frac{4}{3} + \frac{5\sqrt{2}}{3} \right) \frac{\beta_{2}^{2}}{2} . 
\]

iv) For the excited state $ 2P_{1/2} $ with $ n = 2,\: j = 1/2, \: \sigma = - $ ($ n_{r} = 1 $, $ \kappa_{j} = 1 $) in the same approximation $ \mathcal{N}_{2,1/2} \approx 2 $, $ \beta_{2,1/2} \approx \beta_{2} $ and with account of  
\[
2 - \mathcal{N}_{2,1/2} \simeq \frac{\alpha^{2}}{4} , \quad 
K_{1,1/2}^{-}(\rho ) \simeq \frac{2\left(1+2\gamma_{1} \right) }{1 + \varepsilon_{2,1/2}} \left(\ \frac{Z^{2}\alpha^{2}}{4} + \frac{Z^{2}\alpha^{2}}{12} \rho 
+ \frac{1}{9} \rho^{2} \right) ,
\] 
and the diagonal matrix element for $ 2P_{1/2} $ state is  
\[
V_{vac}(2,1/2,-) = -\frac{mc^{2}\alpha^{5}}{48\pi } \left(1+ \frac{3}{8} \alpha^{2}\right) .
\]

\vspace{5mm}

\subsection{\bf{Modification of the Coulomb field by the electron magnetic form factor}}\label{MFF}

Diagonal matrix elements of the perturbation $  V_{mag} ( \mathbf{r}) = -e\delta \varphi_{mag} (r) $ due to the magnetic form factor (\ref{deltafi_magn}) are determined by the integral with the polynomial (\ref{K^2_n})
\[
V_{mag}(n,j,m_{j},\sigma) 
= \lambda_{n,j} A_{n,j}^{2}   
\int_{0}^{\infty}  V_{mag}^{(2)}  \left( \rho \right) e^{-\rho } \rho^{2\gamma_{j}} K_{n_{r},j,\sigma}^{(2)}(\rho ) d\rho  d\rho  .
\]
According to Egs. (\ref{deltafi_magn}), (\ref{fi}) and definition (\ref{rho})
\[
V_{mag}^{(2)}  \left( \rho \right) = \frac{mc^{2} \alpha^{4}}{\pi \mathcal{N}_{n,j}^{2}}  \frac{d\tilde{\phi}}{d\rho} , 
\]
where 
\begin{equation}
	\label{tildafi} 
	\tilde{\phi} \left(\rho \right) = 
	\frac{1}{\rho} \left( 1 - \int_{1}^{\infty} e^{-\frac{\mathcal{N}_{n,j}}{\alpha}\zeta \rho} \frac{d\zeta}{\zeta^{2} \sqrt{\zeta^{2} - 1}} \right). 
\end{equation} 

Therefore, 
\[
V_{mag}(n,j,\sigma) = - \frac{2mc^{2} \alpha^{5}}{\pi \mathcal{N}_{n,j}^{3}} 
\frac{ \left( \mathcal{N}_{n,j} + \kappa_{j} \right) n_{r}! }{4\mathcal{N}_{n,j} \Gamma \left( n_{r} + 1 + 2\gamma_{j} \right)} 
\sum_{\nu = 0}^{2n_{r}} a_{\nu}^{(2)}(n,j,\sigma) \int_{0}^{\infty} \frac{d\tilde{\phi}}{d\rho} e^{-\rho } \rho^{2\gamma_{j} + \nu} d\rho  .  
\] 

Here the integral can be calculated using the relation
\[
\int_{0}^{\infty} \frac{d\tilde{\phi}\left(\rho \right)}{d\rho} e^{-\rho } \rho^{2\gamma_{j} + \nu} d\rho 
= - \int_{0}^{\infty} \tilde{\phi}\left(\rho \right) \frac{d}{d\rho} e^{-\rho } \rho^{2\gamma_{j} + \nu}  d\rho 
\] 
which with account of the explicit expression of $ \tilde{\phi}\left(\rho \right) $ allows to integrate the result over $ \rho $. Thus, we get the expression for the diagonal matrix elements 
\[
V_{mag}(n,j,m_{j},\sigma) = 
= \frac{2mc^{2} \alpha^{5}}{\pi \mathcal{N}_{n,j}^{3}} 
\frac{ \left( \mathcal{N}_{n,j} + \kappa_{j} \right) n_{r}! }{4\mathcal{N}_{n,j} \Gamma \left( n_{r} + 1 + 2\gamma_{j} \right)} 
\sum_{\nu = 0}^{2n_{r}} a_{\nu}^{(2)}(n,j,\sigma) \Gamma \left( 2\gamma_{j} + \nu -1 \right) \times   
\] 
\[
\times \left[ 1 - \left( 2\gamma_{j} + \nu \right) C^{(mag)}_{2\gamma_{j} + \nu -1}  \left( \frac{\alpha}{\mathcal{N}_{n,j}} \right)^{2\gamma_{j} + \nu -1} + \left( 2\gamma_{j} + \nu -1 \right)  C^{(mag)}_{2\gamma_{j} + \nu} \left( \frac{\alpha}{\mathcal{N}_{n,j}} \right)^{2\gamma_{j} + \nu} \right] ,   
\] 
where 
\[
C^{(mag)}_{\mu} = \int_{1}^{\infty} \frac{d\zeta}
{\left( \zeta + \frac{\alpha}{\mathcal{N}_{n,j}} \right)^{\mu}\zeta^{2} \sqrt{\zeta^{2} - 1}} \approx \int_{1}^{\infty} \frac{d\zeta}
{\zeta^{2 + \mu} \sqrt{\zeta^{2} - 1}} .
\]

For the states with the maximal possible value $ j $ at the given $ n $ when $ n_{r} = 0 $, $ \mathcal{N}_{n,j} = n $ and $ P_{0,j}^{(+)} W_{0,j}^{(-)} = 1 $, the matrix element $ V_{mag}(n,+)  $ in approximation $ \gamma_{n} \simeq n $ is  
\[
V_{mag}(n,j_{n},+) = \frac{mc^{2} \alpha^{5}}{2\pi n^{4}\left( 2n-1 \right)}  
\left(  1 - \left( 2n \right) C^{(mag)}_{2n -1} \left( \frac{\alpha}{n} \right)^{2n -1} + \left( 2n -1 \right) C^{(mag)}_{2n}\left( \frac{\alpha}{n} \right)^{2n}  \right)  ,   
\] 

As before, below we calculate the matrix elements for the lowest energy states in the approximation $ \gamma_{1} \simeq 1 $. 

i) For the ground state $ 1S_{1/2} $ with quantum numbers $ n = 1, \: j=1/2, \: \sigma = + $  ($ n_{r} = 0 $, $ \kappa_{1/2} = 1 $, $ \gamma_{1/2} \simeq 1 $), and  
\[
\mathcal{N}_{1,1/2} = 1 , \quad a^{(2)}_{0} = 1 , 
\]
diagonal matrix element is 
\[
V_{mag}(1,1/2,+)= \frac{mc^{2} \alpha^{5}}{2\pi } \left( 1 - \frac{\pi}{2} \alpha + \frac{2}{3}\alpha^{2}  \right)  ,   
\] 

ii) For the excited state $ 2P_{3/2} $ with $ n = 2,\: j = 3/2, \: \sigma = + $ ($ n_{r} = 0 $, $ \kappa_{3/2} = 2 $, $ \gamma_{3/2} \simeq 2 $) and   
\[
\mathcal{N}_{2,3/2} = 2 , \quad a^{(1)}_{0}(2,1/2) = 1 , 
\]
in the same approximation the diagonal matrix element is 
\[
V_{mag}(2,3/2,+) = \frac{mc^{2} \alpha^{5}}{96\pi } \left( 1 - \frac{3\pi}{32} \alpha^{3} + \frac{3}{16} C_{4} \alpha^{4} \right)  ,   
\] 

iii) For the excited state $ 2S_{1/2} $ with $ n = 2,\: j = 1/2, \: \sigma = + $ ($ n_{r} = 1 $, $ \kappa_{j} = 1 $ and 
\[
K^{(2)}_{1,1/2,+}(\rho ) \approx  8 - 6 \rho + \rho^{2}  , \quad  a_{0}^{(2)}(2,1/2,+) = 8 , \quad  a_{1}^{(2)}(2,1/2,+) = - 6 , \quad  a_{2}^{(2)}(2,1/2,+) = 1 . 
\] 
In approximation $ \mathcal{N}_{2,1/2} \simeq 2 $ and neglecting higher orders of $ \alpha $, the matrix element for the $ 2S_{1/2} $ state is 
\[
V_{mag}(2,1/2,+) \simeq \frac{mc^{2} \alpha^{5}}{6\pi }  \frac{3}{8}
\left(1- \frac{\pi}{2} \alpha   
 \right)  ,   
\]

iv) For the excited state $ 2P_{1/2} $ with $ n = 2,\: j = 1/2, \: \sigma = - $ ($ n_{r} = 1 $, $ \kappa_{j} = 1 $) in the same approximation $ \mathcal{N}_{2,1/2} \approx 2 $, $ 2 - \mathcal{N}_{2,1/2} \simeq \alpha^{2}/4 $, and with account of  
\[
K_{(2)}^{-}(\rho ) = \left( \mathcal{N}_{2,1/2}^{2} - 1 \right) \mathcal{N}_{2,1/2} \left( \mathcal{N}_{2,1/2} - 2 \right) - 2 \left( \mathcal{N}_{2,1/2} - 1 \right)^{2} \rho + \frac{\mathcal{N}_{2,1/2} - 1}{\mathcal{N}_{2,1/2} + 1} \rho^{2} \approx 
\] 
\[
\simeq - \frac{3}{2} \alpha^{2} - 2\rho + \frac{1}{3} \rho^{2}
\] 
the diagonal matrix element for $ 2P_{1/2} $ state is  
\[
V_{mag}(2,1/2,-) = -\frac{mc^{2} \alpha^{5}}{48\pi } \left( 1+ \frac{3}{8} \alpha^{2}   \right) . 
\]

\section{The Lamb shift of the $ 2S_{1/2} $ and $ 2P_{1/2} $ hydrogen levels}\label{Lamb-shift}

 In QED the zero-order DE with a Coulomb source 
\[
\left( c \bm{\hat{\alpha}} \cdot \hat{\mathbf{p}} + m c^{2} \hat{\beta} - \frac{e^{2}}{r} \right)\Psi = E \Psi
\] 
provides only an approximate description of hydrogen-like bound states. It  gives the energy spectrum with bound state  levels (\ref{E_n,j}) which depend on the principal quantum number $ n $ and the total angular momentum $ j $, only,  and are degenerate with respect  to number $ \sigma $. This degeneracy is accidental in the sense that it occurs only if the interaction between the electron and proton is exactly proportional to $ \sim 1/r $ as predicted by the Coulomb’s law. Such degeneracy is explained by the  existence in the Coulomb field of the additional Johnson-Lippman integral of motion, non- commuting with the Dirac invariant.

In 1947  in the Lamb–Rutherford experiment \cite{Lamb} a difference between the energy levels of the $ 2S_{1/2} $ and $ 2P_{1/2} $ states of a hydrogen atom was established. Splitting between these levels, denoted as the Lamb shift, removes the degeneracy in the spectrum of the DE in the Coulomb field. The very existence of the Lamb shift indicates that the  Coulomb’s law fails at short distance scales near the atomic nucleus. In this paper we have considered the main factors which break the Coulomb symmetry, namely, modification of the Coulomb’s law predicted in QED.

First of all, QED is the theory of the interacting spinor (Dirac) and electromagnetic (photon) fields in which electrically charged particles interact by means of the exchange by photons, and photon propagator describes this interaction. Photon interaction with the electron-positron spinor field inserts  the polarization operator, related to virtual emission and absorption of electron-positron pairs, in the photon propagator. This effect leads to modification of the Coulomb potential in the form of the Uehling potential (\ref{RadCorr}). 

Secondly, an electron continuously emits and absorbs virtual photons and as a result, its electric charge is spread over a finite volume described by electron form factors in (\ref{A0e}). This also leads to effective modifications (\ref{Coul_pert})-(\ref{PP}) of the electron interaction with external charge. The electric form factor modifies the Coulomb law and the magnetic form factor describes radiative corrections to the spin-orbit coupling. 

Thirdly, we take into account also the vacuum polarization of the external charge by the electric field. At strong external electric fields $ \mathcal{E} \sim \pi m^{2} c^{3}/e\hbar $ the creation of real electron-positron pairs from vacuum  is possible (so called Schwinger effect). It is easy to estimate that for a point charge such fields correspond to distance $ r^{2} = (\alpha /\pi) \lambda_{C}^{2} $ ($\alpha=e^2/\hbar c$ is the fine structure constant and $\lambda _c=h/mc $ is the Compton wavelength). The vacuum polarization, which was considered for the  first time by W. Heisenberg and H. Euler, prevents such processes. To our knowledge, the influence of the  vacuum polarization by the  electric field of an external charge on the Lamb shift was not considered previously.  

In this paper the solution (\ref{bisp_sigma})-(\ref{E_n,j}) of the DE with the Coulomb potential was used as a starting point to obtain the bound energy spectrum of hydrogen atom when all three above mentioned effects contribute to the Lamb shift. As a result, the energy spectrum of the hydrogen atom is described by the Hamiltonian 
\[
\mathrm{H} = \sum_{n,j,m_{j},\sigma} \tilde{E}_{n,j,\sigma} a_{n,j,m_{j},\sigma}^{\dagger} a_{n,j,m_{j},\sigma}  
\] 
where 
\[ 
\tilde{E}_{n,j,\sigma} = mc^{2} \varepsilon_{n,j} + \Delta_{n,j,\sigma}.
\] 
Here the value
\[
\Delta_{n,j,\sigma} \equiv V_{PP}\left( n,j,\sigma \right) + V_{elec}\left( n,j,\sigma \right) + V_{vac}(n,j,\sigma) + V_{mag}(n,j,\sigma)
\]
determines the Lamb shifts of the levels. Although initially the Lamb shift was defined as the splitting between $ 2S_{1/2} $ and $ 2P_{1/2} $ states, the difference in levels $ n \neq 2 $ are also referred to as Lamb shifts. In particular, for  $\Delta_L\equiv \Delta_{2S_{1/2}} - \Delta_{2P_{1/2}} $ we have
\[\Delta_L 
\simeq  \frac{mc^{2}\alpha^{5}}{6\pi } \left[ \ln \frac{2m}{\lambda} + \frac{1}{8} - \frac{1}{5}  
- \frac{3}{8} \left( \frac{5\pi}{8} \ln \frac{4m}{\lambda} - \frac{11\pi}{16} \right) \alpha  + \right.  
\] 
\[
\left. +  \frac{1}{4} \sqrt{\frac{\pi \alpha}{30}} \left( 
\frac{8}{3} \mathbb{C}  
+ \frac{8}{3} \ln \frac{\sqrt{2}}{\beta_{2}} 
+ \frac{4}{3}\sqrt{2} \ln \frac{\sqrt{2} + 1}{\sqrt{2} - 1} 
+ \frac{5}{3} \ln 2 
+ \frac{\pi}{3} 
+ \frac{22}{9} + \frac{10\sqrt{2}}{9} \right) \right] . 
\]

Recall, the expression for the electric form factor $ F_{1} (q^{2} ) $ in Eq. (\ref{vert-corr}) contains in the logarithm argument the parameter $ \lambda \rightarrow 0 $ which is "finite virtual photon mass" introduced to remove the infrared singularity. It plays the role of a low-frequency cutoff parameter. In the standard calculations, this parameter in the general case of arbitrary $Z$ is assumed to be in the interval $ \left( Z\alpha \right)^{2}m \ll \lambda \ll m $ and after addition of the low-frequency contribution parameter, $ \lambda $ cancels out. Minimization of the low-frequency contribution leads to the equality $ \ln \left( 2m /\lambda \right) \rightarrow \ln \left( 1/Z^{2}\alpha^{2} \right) + const $, where  $ const \sim 1 $ \cite{RelQuant,Akhiezer,Flambaum}, which allows to use it as some kind of the fitting parameter. 

The effective potentials (\ref{deltafi_elec}) and (\ref{deltafi_magn}) were first considered in \cite{Flambaum} and used in calculations of the  relativistic heavy elements and in molecular calculations (see \cite{Ginges,Sunaga}). As above, here we consider the hydrogen spectrum ($Z=1$) and according to Schwinger's analysis \cite{Schw}, we choose
\begin{equation}
	\label{lambda_ef2} 
	\ln \frac{2m}{\lambda} \rightarrow \ln {\frac{1}{\alpha^{2}}} - 2.8118  
\end{equation} 
instead of Eq. (\ref{lambda_ef}). Julian Schwinger used the following numerical values 
\[
\alpha = \frac{1}{137.06} , \quad \ln \frac{m}{\vert E_{1}  \vert} = \ln \frac{2}{\alpha^{2}} \approx 10.5 , \quad \frac{\alpha^{3}}{3\pi} = \frac{mc^{2}\alpha^{5}}{6\pi} = 135.644 MHz 
\] 
and for the electron mass he used the equivalent mass (factor $ M_{p}/(M_{p} + m_{e}) $) which gives the  value 
\[
\Delta_{L} = E_{2S_{1/2}} - E_{2P_{1/2}} = 1050.55 MHz .
\]  

At present the following values are accepted 
\[
\alpha = \frac{1}{137.036} , \quad \Delta^{(theor)}_{L} = 1057.864 MHz , \quad \Delta^{(exp)}_{L} = 1057.845 MHz .
\]
For our estimation, the Lamb shift $ \Delta_{L} $ can be represented as 
\[
\Delta_{L} = \frac{mc^{2}\alpha^{5}}{6\pi } C_{L}
\]
where 
\[
C_{L} = \left[ \ln \frac{2m}{\lambda} + \frac{1}{8} - \frac{1}{5}  
- \frac{3}{8} \left( \frac{5\pi}{8} \ln \frac{4m}{\lambda} - \frac{11\pi}{16} \right) \alpha  + \right.  
\] 
\[
\left. +  \frac{1}{4} \sqrt{\frac{\pi \alpha}{30}} \left( 
\frac{8}{3} \mathbb{C}  
+ \frac{8}{3} \ln \frac{\sqrt{2}}{\beta_{2}} 
+ \frac{4}{3}\sqrt{2} \ln \frac{\sqrt{2} + 1}{\sqrt{2} - 1} 
+ \frac{5}{3} \ln 2 
+ \frac{\pi}{3} 
+ \frac{22}{9} + \frac{10\sqrt{2}}{9} \right) \right] . 
\]
Here constant $C_L$ can be represented as $C_L=  \ln \frac{2m}{\lambda}+D$, where $D=0.400759$. Estimating now $ C_{L} $ from  $ \Delta^{(exp)}_{L} $, we can calculate the value $ \ln (2m/\lambda) $ and estimate the difference 
\[
\ln \frac{2m}{\lambda} - \ln {\frac{1}{\alpha^{2}}} \approx -2.44262, 
\]
which is very close to $ \sim - 2.8118 $, and, therefore, we have a good argument in favor of  our calculations.  

\vskip5mm 

\section{Conclusions}

Taking into account that the correct description of a hydrogen atom can only be ensured by the DE, we have calculated the bound electron states of a hydrogen atom with account of  the   Coulomb law modification in the vicinity of the atomic nucleus predicted by QED. The deviation of the electric field from the Coulomb law is the main reason which removes the "accidental" degeneracy of the hydrogen-like energy levels and results in the Lamb shift. The DE and Lamb shift are discussed in any textbook on quantum electrodynamics and quantum field theory (see.e.g.,  \cite{RelQuant,Akhiezer,Gusynin}). The early results on the Lamb shift can be found in the classical book by Bethe and Salpeter \cite{Bethe} and the modern state of the theory is given in the comprehensive review \cite{Eides}. 

In the field-theoretical approach, calculations of the corrections to the energy levels are based on the covariant form of the QED Lagrange functional. Electrodynamic corrections are searched in the form of the power series expansion over small parameter $ \alpha $  \cite{RelQuant,Akhiezer,Gusynin}. Within this scheme the Lamb shift is traditionally obtained using the perturbation theory in the form of the power series expansion up to the given power of $ \alpha $,  and the eigen functions of the Schr\"odinger equation as the non-relativistic approximation of the DE are always used to calculate physical quantities.

Our idea is based on the DE and is principally different from this scheme. We derived the relativistic Hamiltonian given in Eq. (\ref{H_DA}) of the hydrogen atom with external fields using the solution of the DE with a Coulomb source as a starting point. As a perturbation in this Hamiltonian, we have considered the deviation of the electric field, which acts on the electron, from the Coulomb law, and calculated the Lamb shift within the conventional in quantum mechanics Hamiltonian description. The perturbation matrix elements $ V_{\lbrace \nu \rbrace ,\lbrace \nu' \rbrace } $ are determined by the exact solution of the DE with the Coulomb potential. Therefore, the relativistic Hamiltonian in the form (\ref{H_DA}) contains all the so called \textit{relativistic}
or \textit{binding corrections} \cite{Eides}. This approach seems to be more consequent as comparing with other calculations (e.g., diagrams). 

Worth mentioning, here we have taken into account only the RCs that modify the Coulomb law near a nucleus which results in the Lamb splitting between various energy levels in the general case. In the   case of the lowest hydrogen states $ 2S_{1/2} $ and $ 2P_{1/2} $ the Lamb shift equals, according to our calculations, 1,340 MHz, neglecting terms $ \sim \alpha^{2} $.

 A novel feature of our approach is the consistent calculation of the vacuum polarization input into the energy levels performed in Subsection \ref{vac-pol}.  So far to our knowledge, it is the first calculation of such a kind. Usually in this context the vacuum polarization is treated as the factor which leads to the modification of the Coulomb law in the form of the Uehling potential. 
  Moreover, as has been shown above, the vacuum polarization by the proton charge is automatically (i.e., without any additional assumption) accompanied  by an appearance of 
  some spatial scale $r_0$ (see Eq. (\ref{r_0})) which, according to Eqs. (\ref{Er}) and (\ref{dvac}), characterizes qualitatively different spatial behavior of the Coulomb law at $r\ll r_0$ and $r\gg r_0$ and can be considered as the vacuum nuclear polarization length. It can be indirectly associated with the  proton radius, though has another value.

  As the simplest of all stable atoms, hydrogen is unique in the sense of its usefulness for comparison of theory and experiments on bound-state energy level structures. Nowadays precise optical spectroscopy and theoretical calculations \cite{Eides} have improved tremendously and reached a point where the proton size is the limiting factor when comparing experiment with theory. Although the energy levels shifts  associated with the proton finite size, are small, the root-mean-square charge radius (rms radius) can be determined based on the high precision spectroscopy and QED calculations of the bound-states. The Lamb shift is dominated by purely radiative effects and to extract a value of proton radii $ r_{p} $ from spectroscopic data, it is necessary to minimize theoretical uncertainty in QED radiative corrections. In \cite{Pohl,Antognini} $ r_{p} $ was determined from the spectroscopy of muonic hydrogen ($ \mu p $, that is, a proton orbited by a muon). Obtained from the measurement of a muonic Lamb shift and on the basis of present QED calculations, proton radius differs from determined by electron–proton scattering experiments and from the CODATA value \cite{Mohr}, and the authors concluded that origin of the discrepancy with the data could originate from the wrong or missing QED terms or from unexpectedly large contributions of yet not taken into account higher order terms. 
  
  At present, theoretical value for the $ 2S_{1/2} - 2P_{1/2} $ energy difference 1057.833 MHz is very close to the best experimental value 1057.845 MHz which is an outstanding evidence of QED validity. The difference can be connected with the radius $ r_{0} $, defined in Eq. (\ref{r_0}),  due to the vacuum polarization by nucleus charge considered in the present paper, which has not been accounted previously. The exact solution of the DE with the Coulomb potential used to calculate matrix elements, gives an account of higher order terms of $ \alpha $. 
  
  Here we did not aim to get the expressions for the Lamb shift with the high order accuracy with respect to the constant $\alpha $ because, as it was mention above, considered here RCs are only part of electrodynamic effects which contribute to the Lamb shift. Therefore, a quantitative estimation of the Lamb shift can not be taken without account of a proton magnetic moment (hyperfine structure) and recoil effect (electron-proton reduced mass corrections). Account of the corresponding effects demands separate study.

\vskip5mm 
{\bf Acknowledgement.} 
\textit{ This work was supported by the Department of Physics and Astronomy of the National Academy of Sciences of Ukraine (fundamental scientific program 0122U000887) and the Simons Foundation (USA).}

\vskip5mm 

\appendix 
\section{Radial functions and spherical spinors} \label{RadF-SphSpin}
\subsection{Radial functions}\label{RadF}

The explicit expressions of the radial functions are 
\[
R_{n,j}^{(\sigma)} (\rho ) = R_{n_{r},j} (\rho ) P_{n_{r},j}^{(\sigma)} (\rho ), \]
 \begin{equation}
	\label{radFunc} 
 Q_{n,j}^{(\sigma)} (\rho ) = R_{n_{r},j} (\rho ) W_{n_{r},j}^{(\sigma)} (\rho ) 
\end{equation}
where $ \rho $ was defined in Eq. (\ref{rho}). Thus, we have  
\begin{equation}
\label{R_n,j} 
	R_{n_{r},j} (\rho ) = \left( \frac{2}{r_{B}\mathcal{N}_{n_{r},j}} \right)^{3/2}  A_{n,j} e^{-\rho /2} \rho^{\gamma_{j}-1} , 
\end{equation}
\begin{equation}
	\label{A_n,j} 	
	A_{n,j} = \sqrt{\frac{\left( 1 + \varepsilon_{n_{r},j} \right) \left( \mathcal{N}_{n_{r},j} + \kappa_{j} \right) n_{r}! }{4\mathcal{N}_{n_{r},j} \Gamma \left( n_{r} + 1 + 2\gamma_{j} \right)}}
\end{equation} 
and $ P_{n_{r},j}^{(\pm)} (\rho ) $, $ W_{n_{r},j}^{(\mp)} (\rho) $ are polynomials of the order $ n_{r} $ expressed via the generalized Laguerre polynomials $ \mathit{L}_{n}^{2\gamma}(\rho ) $ 
\begin{equation}
\label{polP,W} 
\begin{array}{c}
P_{n_{r},j}^{(+)} (\rho ) = \mathit{L}_{n_{r}}^{2\gamma_{j}} (\rho) - \frac{n_{r} + 2\gamma_{j}}{\mathcal{N}_{n,j}+\kappa_{j}} \mathit{L}_{n_{r}-1}^{2\gamma_{j}} (\rho) , \\ 
W_{n_{r},j}^{(-)} (\rho ) = \mathit{L}_{n_{r}}^{2\gamma_{j}} (\rho) + \frac{n_{r} + 2\gamma_{j}}{\mathcal{N}_{n,j}+\kappa_{j}} \mathit{L}_{n_{r}-1}^{2\gamma_{j}} (\rho) , \\ 
P_{n_{r},j}^{(-)} (\rho ) = \frac{\sqrt{n_{r}\left( n_{r} + 2\gamma_{j} \right) } }{\mathcal{N}_{n,j}+\kappa_{j}} \mathit{L}_{n_{r}}^{2\gamma_{j}} (\rho) - \sqrt{\frac{n_{r} + 2\gamma_{j}}{n_{r}}} \mathit{L}_{n_{r}-1}^{2\gamma_{j}} (\rho) , \\ 
W_{n_{r},j}^{(+)} (\rho) = \frac{\sqrt{n_{r}\left( n_{r} + 2\gamma_{j} \right) } }{\mathcal{N}_{n,j}+\kappa_{j}} \mathit{L}_{n_{r}}^{2\gamma_{j}} (\rho) + \sqrt{\frac{n_{r} + 2\gamma_{j}}{n_{r}}} \mathit{L}_{n_{r}-1}^{2\gamma_{j}} (\rho) . 
\end{array}
\end{equation}

They  can be easily found from the expression 
\[
\mathit{L}_{n}^{2\gamma} \left( \rho \right) = \frac{1}{n!} e^{\rho} \rho^{-2\gamma} \frac{d^{n}}{d\rho^{n}} \left( e^{-\rho} \rho^{n+2\gamma} \right) = \]
\[\sum_{m=0}^{n} \left( -1 \right)^{m} {\begin{pmatrix}n+2\gamma\\n-m\\\end{pmatrix}} \frac{\rho^{m}}{m!} .
\] 
Several special cases of $ \mathit{L}_{n}^{2\gamma} $ are:
\[
\mathit{L}_{0}^{2\gamma} = 1 , \quad \mathit{L}_{1}^{2\gamma} = 2\gamma + 1 - \rho , \]
\[\mathit{L}_{2}^{2\gamma} = \left( 1 + \gamma \right) \left( 1 + 2\gamma \right) - 2\left( 1 + \gamma \right) \rho + \frac{1}{2} \rho^{2} .
\] 

For states with $ j = n - 1/2 $ when $ \kappa_{n-1/2} = n $, $ n_{r} = 0 $ and $ \mathcal{N}_{n,n-1/2} = n $, the energy levels are 
\begin{equation}
\label{eps_N} 
\varepsilon_{0,n-1/2} \equiv \varepsilon_{n} = \sqrt{1 - \frac{ \alpha^{2}}{n^{2}} } = 
\frac{\gamma_{n}}{n}, 
\end{equation}
 \[\gamma_{n} = \sqrt{n^{2} - \alpha^{2} }\] 
and the polynomials (\ref{polP,W}) of the order $ n_{r} = 0 $ in (\ref{polP,W}) are reduced to constants $ P_{0,j}^{(+)} = W_{0,j}^{(-)} = 1 $ and $ P_{0,j}^{(-)} = W_{0,j}^{(+)} = 0 $. Hence, for these states $ \sigma $ takes one sign $ \sigma = + $, only, and the eigenbispinors are given by the expression
\[
\Psi_{n, n-1/2, m_{j},+ } (\mathbf{r}) = {\begin{pmatrix}\psi_{n,m_{j}}^{(u)}(\mathbf{r})\\ \lambda_{n} \psi_{n,m_{j}}^{(d)}(\mathbf{r})\\\end{pmatrix}} =\]

\begin{equation}
	\label{Bisp_j_n,+} 
	\qquad =
 {\begin{pmatrix}R_{n}(\rho ) \Omega_{n, m_{j},+}\\ i \lambda_{n} R_{n}(\rho ) \hat{\sigma}_{r} \Omega_{n, m_{j},+}\\
\end{pmatrix}} 
\end{equation}
with the radial function $ R_{n} (\rho) = R_{0,j_{n}} $ ($ j_{n} = n - 1/2 $) (see (\ref{R_n,j}))
\begin{equation}
\label{R_(n,j_n)} 
R_{n}(\rho) = \left( \frac{2}{nr_{B}} \right)^{3/2} \sqrt{\frac{ 1 + \varepsilon_{n} }{2 \Gamma \left( 1 + 2\gamma_{n} \right)}} e^{-\frac{\rho}{2 }} \rho^{\gamma_{n}-1}, 
\end{equation}
where in this particular case 
\[ \rho = \frac{2r}{n r_{B}} , \quad \lambda_{n} = \sqrt{\frac{1-\varepsilon_{n}}{1+\varepsilon_{n}}}. 
\]

\subsection{Spherical spinors} \label{Sph-Spin}

The Dirac bispinor (\ref{bisp_sigma}) includes spherical spinors $ \Omega_{j-\sigma (1/2), m_{j},\sigma} $ in which number $ j-\sigma (1/2) = l $ takes the integer values $ l = 0,1,2,\ldots $. There are two spinors with $ \sigma = + $ and $ \sigma = - $ 
\[
\Omega_{l,m_{j},+} = \frac{e^{im_{j}\varphi}}{\sqrt{2\pi}} \left( \begin{array}{c}
\sqrt{\frac{l + 1/2 + m_{j}}{2l + 1}} e^{-i\varphi /2} \mathcal{P}_{l}^{ m_{j}-1/2 } \\
\sqrt{\frac{l+1/2 - m_{j}}{2l+1}} e^{i\varphi /2} \mathcal{P}_{l}^{ m_{j}+1/2 } 
\end{array} \right) =\]
\begin{equation}
\label{Omega_+}  
\qquad	= \left( \begin{array}{c}
\sqrt{\frac{l + 1/2 + m_{j}}{2l + 1}} Y_{lm_{1}} (\vartheta ,\varphi) \\
\sqrt{\frac{l+1/2 - m_{j}}{2l+1}} Y_{lm_{2}} (\vartheta ,\varphi) 
\end{array} \right), 
\end{equation}
\[ \Omega_{l,m_{j},-} = \frac{e^{im_{j}\varphi}}{\sqrt{2\pi}} \left( \begin{array}{c}
\sqrt{\frac{l + 1/2 - m_{j}}{2l + 1}} e^{-i\varphi /2} \mathcal{P}_{l}^{ m_{j}-1/2 } \\
-\sqrt{\frac{l+1/2 + m_{j}}{2l+1}} e^{i\varphi /2} \mathcal{P}_{l}^{ m_{j}+1/2 } 
\end{array} \right) = \]
\begin{equation}
	\label{Omega_-} 
\qquad = \left( \begin{array}{c}
\sqrt{\frac{l + 1/2 - m_{j}}{2l + 1}} Y_{lm_{1}} (\vartheta ,\varphi) \\
-\sqrt{\frac{l+1/2 + m_{j}}{2l+1}} Y_{lm_{2}} (\vartheta ,\varphi) 
\end{array} \right).
\end{equation}
which are the eigen spinors of the matrix $ \hat{\Lambda} $ corresponding to two eigenvalues. One of them  is positive, $ \lambda_{1} = \lambda_{+}(l) = \hbar (l + 1) $ for $ \Omega_{l,m_{j},+} $ with $ l = l_{1} = j - 1/2 $, and another one is negative, $ \lambda_{2} = \lambda_{-}(l) = -\hbar l $ for $ \Omega_{l,m_{j},-} $ with $ l = l_{2} = j + 1/2 $. 

In Eqs. (\ref{Omega_+})-(\ref{Omega_-}) functions $ Y_{lm} $ are the normalized spherical harmonic functions 
\[
Y_{lm} (\vartheta ,\varphi) = \frac{1}{\sqrt{2\pi}} e^{im\varphi} \mathcal{P}_{l}^{ m } \left( \vartheta \right) 
\] 
with $ m = m_{1} = m_{j}-1/2 $ and $ m = m_{2} = m_{j}+1/2 $ ($ m_{2} - m_{1} = 1 $). The polynomials $ \mathcal{P}_{l}^{ m } \left( \vartheta \right) $ in $ Y_{lm} $ are the normalized associated Legendre polynomials

\begin{equation}
\label{spherharm} 
\mathcal{P}_{l}^{ m } \left( \cos \vartheta \right)=
\end{equation}
\[
= i^{l} \left( -1 \right)^{\frac{m+\vert m \vert}{2}} \sqrt{\frac{2l+1}{2} \frac{\left( l-\vert m \vert \right)!}{\left( l+\vert m \vert \right)!} } P^{\vert m \vert}_{l} \left( \cos \vartheta \right) 
\]
which provide the normalization condition: 
\[
\int Y_{lm}^{\ast} \left( \mathbf{e}_{\mathbf{r}}\right) Y_{l'm'} \left( \mathbf{e}_{\mathbf{r}}\right) do = \int_{0}^{2\pi} d\varphi \int_{0}^{\pi} d\vartheta \sin \vartheta \times \]
\begin{equation}
	\label{normcond} 
\quad \times  Y_{lm}^{\ast} (\vartheta ,\varphi) Y_{l'm'} (\vartheta ,\varphi) = \delta_{l,l'} \delta_{m,m'}.
\end{equation} 

Each eigen bispinor of the DE includes two spherical spinors with different numbers $ l $,  namely, $ l_{1} = j - 1/2 $ and $ l_{2} = j + 1/2 $. These numbers take integer values  $ l = 0,1,2,\ldots $ and indicate the degree of the corresponding associated Legendre polynomial. Spherical spinors are connected by the following relations 
\[
%\begin{array}{c}
\hat{\sigma}_{r} \Omega_{j-1/2, m_{j},+} = i \Omega_{j+1/2, m_{j},-}, \] \[\hat{\sigma}_{r} \Omega_{j+1/2, m_{j},-} = -i \Omega_{j-1/2, m_{j},+},\] 
or
\[
 \hat{\sigma}_{r} \Omega_{j-\sigma 1/2, m_{j},\sigma} = i\sigma \Omega_{j+\sigma 1/2, m_{j},-\sigma}, \; \sigma = \pm . 
%\end{array} 
\]
where (see (\ref{PM})) $ \hat{\sigma}_{r} = \bm{\hat{\sigma}} \cdot \mathbf{e}_{r} $ and vector $ \mathbf{e}_{r} = \mathbf{r} /r $ is the unit vector in the direction of $ \mathbf{r} $ .

\end{document}